\documentstyle[aps,epsf]{revtex}
\draft
\begin{document}
\title{Tunneling and Traversal of Ultra-cold Atoms through Vacuum Induced Potentials}
\author{R. Arun$^{1}$ and G. S. Agarwal$^{1,2}$}
\address{$^{1}$Physical Research Laboratory, Navrangpura, Ahmedabad 380 009,
India}
\address{$^{2}$Jawaharlal Nehru Centre for Advanced Scientific Research,
Bangalore, India}
\date{\today}
\maketitle
\begin{abstract}
We examine the passage of ultra-cold two-level atoms through the potential
produced by the vacuum of the cavity field. The peak of the transmitted wave packet generally 
occurs at the instant given by the expression for phase time even if that instant is
earlier than the instant at which incident wave packet's peak enters the cavity and thus
the phase time can be considered as the appropriate measure of the time required
for the atom to traverse the cavity. We show that the phase tunneling time for
ultra-cold atoms could be both super - and sub - classical time and we show how
this behaviour can be understood in terms of the momentum dependence of the phase  
of transmission amplitude. The passage of the atom through the cavity is unique as it
involves a coherent addition of the transition amplitudes corresponding to both
barrier and well. New features such as the splitting of the wave packet arise
from the entanglement between the center of mass motion and the electronic as
well as field states in the cavity.
\end{abstract}

\newpage
\section{introduction}
An important question of great interest in several disciplines of physics
has been - what is the tunneling time or traversal time of a quantum
mechanical particle through a potential. Various definitions have been
proposed and the subject has been reviewed extensively
\cite{{review},{hartmann},{chiao},{mexican},{dwell}}.
Although there exists no unique definition
for the tunneling time, an important experiment by Steinberg et al \cite{expt1}
supports the belief that tunneling time is consistent with the Wigner's phase
time \cite{wigner} defined as the frequency derivative of the transmission
amplitude phase. Steinberg et al observed that tunneling velocity of a single photon
wave packet passing through the forbidden midgap region of a photonic
band-gap meterial is superluminal.

In this paper, we examine the passage of a cold atom through a high quality
cavity. In particular, we enquire what is the passage time of the atom
through the cavity. The question is a complicated one as we have here a
coupling with three different types of the degrees of freedom - (a) atom's
center of mass motion, (b) atom's electronic states and (c) photons. Besides
in a high quality cavity, the resonant transitions are especially important.
Thus in the passage through the cavity the atom can change its electronic
state. We have found that the passage time can be defined through the
complex transmission amplitude. The connection to potential problems is
provided by the existing results in the context of micromazers 
\cite{{scully},{meyer}}. An analysis in the
the dressed state basis reveals that the interaction of a moving atom with a 
single mode vacuum field in a high quality cavity is equivalent to a combination 
of a potential barrier and a well. These
potentials belong to the category of vacuum induced potentials and should
be distinguished from the optical potentials produced by a far off
resonant field interacting with an atom \cite{optical}. Having realized 
that the cavity field can act like a potential for an ultra-cold atom,
one could calculate the time the atom takes to traverse the cavity
using methods similar to those used say in the context of tunneling 
electrons through potential barriers.

The organization of the paper is as follows. In Sec. II, we formulate
the model describing the propagation of ultra-cold atoms through a
high quality cavity. In Sec. III, we calculate the phase time for
ultra-cold atoms. We find that the phase time could also be negative,
which is reminiscent of superluminal propagation of electromagnetic
fields\cite{wang}. We explain this characteristic of phase time in terms 
of the dispersion of the phase of transmission amplitude. This is
again very much analogous to the propagation of light which can be
understood in terms of the dispersion of the medium. In Sec. IV, we
present the numerical results for the time dependence of the wave
packet. We show the correlation of the peaks of wave packets to the
phase time of Sec. III. In Sec. V, we discuss aspects of phase time  
where the effects of the entanglement with the internal degrees of
freedom is important. Finally, we summarize our results in Sec. VI.
 
\section{model system and summary of atom - field interaction}
We consider an ultra-cold, two-level atom in its excited state to be
incident on a single mode cavity of length $L$ as shown in Fig. 1.
The frequency of the cavity field has been tuned to the frequency 
of the atomic transition between the excited state $|e\rangle$ and
the ground state $|g\rangle$. The Hamiltonian describing this
resonant atom-field interaction including the quantized motion of
center-of-mass (c.m.) of the atom, is given by 
\begin{equation}
H = \frac{p_z^2}{2 m} + \hbar \omega_e~|e \rangle \langle e| + 
\hbar \omega_g~|g \rangle \langle g| + \hbar \omega~a^{\dagger} a 
+ \hbar g~u(z) (\sigma a^{\dagger} + a \sigma^{\dagger})~,
\label{hamilton1} 
\end{equation}
where $g$ is the atom-field coupling constant and $\omega =
\omega_e - \omega_g$ is the resonance frequency of atom-field
interaction. The operators $a~(a^{\dagger})$ annihilate (create)
a photon of frequency $\omega$ and $\sigma~(\sigma^{\dagger})$ are
the lowering (raising) operator for the atomic transition. For
simplicity, we approximate the mode function $u(z)$ of the cavity 
by a mesa function. 

In a reference frame rotating with frequency $\omega$ the Hamiltonian
of the atom-field interaction reads as
\begin{equation}
H_I = \frac{p_z^2}{2 m} + \hbar g~u(z) (\sigma a^{\dagger} + 
a \sigma^{\dagger})~.
\label{hamilton2}
\end{equation}
The operator $(\sigma a^{\dagger} + a \sigma^{\dagger})$ is easily 
diagonalizable. It has eigenstates $|\phi^{0}\rangle$, 
$|\phi_{n+1}^{\pm}\rangle$ with eigenvalues $0$, $\pm \sqrt{n + 1}$,
respectively. The dressed eigenstates can be expanded in terms of
eigenstates of the free Hamiltonian as $|\phi^{0}\rangle = |g,0\rangle$  
and $|\phi_{n+1}^{\pm}\rangle = \frac{1}{\sqrt{2}}(|e,n\rangle \pm
|g,n+1 \rangle)$. If we expand the combined state of atom-cavity 
system as 
\begin{equation}
|\Psi \rangle = \chi_{0}(z,t) |\phi^{0}\rangle + 
\chi_{+}(z,t) |\phi_{n+1}^{+}\rangle + \chi_{-}(z,t) |\phi_{n+1}^{-}\rangle~,
\label{total}
\end{equation}
then the time dependent Schrodinger equation becomes 
\begin{equation}
i \hbar \frac{\partial \chi_{\alpha}}{\partial t} = h_{\alpha} \chi_{\alpha}~,
~~~~\alpha = \pm,0~.
\label{time}
\end{equation} 
Here $h_o = p_z^2/2 m$, $h_{\pm} = p_z^2/2 m 
\pm \hbar g u(z) \sqrt{n + 1}$ are operators acting in the space of the
center of mass variables. Clearly the cavity with fixed number of 
photons creates a barrier and a well potential for the external motion of
the atom corresponding to the dressed states $|\phi_{n+1}^{\pm}\rangle$
respectively as discussed in Ref. \cite{scully}. For the mesa mode
function $u(z) = \theta(z) \theta(L - z)$, the cavity induced potentials
are displayed in Fig. (2).      

Since we need the transmission amplitude of the excited atom for further 
discussion we summarize the main results of Meyer et al \cite{meyer}.
Consider the initial atom-field state to be $|e,n\rangle$, i.e., 
the atom is in the excited state and the cavity field contains $n$  
photons. This initial atom-field state can be expanded in terms of
dressed states as 
\begin{equation}
|e,n\rangle = \frac{1}{\sqrt{2}} \left[|\phi_{n+1}^{+}\rangle + 
|\phi_{n+1}^{-}\rangle \right]~.
\label{expand}
\end{equation}
From the above discussions, it is clear that the external motion of
the excited atom experiences a coherent addition of a potential barrier 
and a well with potential energy $V = \hbar g \sqrt{n + 1}$. We expand
the initial wavepacket of the atom incident on the cavity 
as $\psi(z,0) = \int dk A(k) e^{ikz} \theta(-z)$. The
Heaviside's step function $\theta(-z)$ merely reflects the fact 
atomic wavepacket enters the cavity from the left side. The fourier
amplitudes $A(k)$ are adjusted such that the center of the wave 
packet enters the cavity at time $t = 0$. The initial wave function
of the atom-field system is therefore
\begin{equation}
| \Psi(z,0) \rangle = \psi(z,0) |e,n\rangle~.
\end{equation}   
The wave function of the atom-field system after the atom has
left the cavity is given by using Eq. $(\ref{expand})$
\begin{eqnarray}
|\Psi(z,t)\rangle &=& \exp\left(\frac{-i H_I t}{\hbar}\right) 
\psi(z,0) |e,n\rangle \nonumber \\ 
&=& \frac{1}{\sqrt{2}} \left[  \exp\left(\frac{-i H_I t}{\hbar}\right) 
|\phi_{n+1}^{+}\rangle \psi(z,0) + \exp\left(\frac{-i H_I t}{\hbar}\right)
|\phi_{n+1}^{-}\rangle \psi(z,0) \right]~.
\label{main1} 
\end{eqnarray}
It is to be noted that the first and second term in the above equation
corresponds to the atom interacting with the potential barrier and well 
respectively. Denoting the reflection and transmission amplitudes 
as $\rho_{n}^{\pm}$, $\tau_{n}^{\pm}$ for the potential barrier 
(superscript +) and well (superscript -) respectively, we have 
\begin{equation}
\rho_{n}^{\pm} = i \Delta_{n}^{\pm} \sin(k_n^{\pm} L) \exp(i k L) \tau_n^{\pm}~,
\label{rhos}
\end{equation}

\begin{equation}
\tau_n^{\pm} = \exp(-i k L){\left[\cos(k_n^{\pm} L) -i \Sigma_n^{\pm}
\sin(k_n^{\pm} L) \right]}^{-1}~,
\label{taus}
\end{equation}

\begin{eqnarray}
\Delta_n^{\pm} &=& \frac{1}{2}\left(\frac{k_n^{\pm}}{k} - \frac{k}{k_n^{\pm}}
\right)~, \nonumber \\
\Sigma_n^{\pm} &=& \frac{1}{2}\left(\frac{k_n^{\pm}}{k} + \frac{k}{k_n^{\pm}}
\right)~, 
\label{sigmas}
\end{eqnarray}

\begin{eqnarray}
k_n^{\pm} &=& \sqrt{\left(k^2 \mp \frac{2mg}{\hbar} \sqrt{n + 1} \right)}
\nonumber \\
&=& \sqrt{\left(k^2 \mp k_o^{2} \sqrt{n + 1} \right) }~.
\label{wave}
\end{eqnarray}
Here $\hbar k$ is the atomic c.m. momentum and $\hbar^2 k_o^{2}/2 m
= \hbar g $ is the vacuum coupling energy. Carrying out the time 
evolution for the dressed states, we get the following wave 
function after the atom - field interaction : 
\begin{eqnarray}
| \Psi(z,t)\rangle &=& \int dk A(k) e^{-i \left(\hbar k^2/2 m \right) t}
\left \{ \left[R_{e,n}(k) e^{-ikz} \theta(-z) + T_{e,n}(k) e^{i k z}
\theta(z-L) \right] |e,n\rangle \right. \nonumber \\
&+& \left. \left[R_{g,n+1}(k) e^{-i k z} \theta(-z) + T_{g,n+1}(k)
e^{i k z} \theta(z-L) \right] |g,n+1 \rangle \right \}~,
\label{main2}
\end{eqnarray}   
where 
\begin{equation}
R_{e,n} = \frac{1}{2}(\rho_n^{+} + \rho_n^{-}),~~~~~T_{e,n} = 
\frac{1}{2}(\tau_n^{+} + \tau_n^{-})~,
\label{excite} 
\end{equation}
are the reflection and transmission amplitudes for the excited state of
the atom and
\begin{equation}
R_{g,n+1} = \frac{1}{2}(\rho_n^{+} - \rho_n^{-}),~~~~~T_{g,n+1} =
\frac{1}{2}(\tau_n^{+} - \tau_n^{-})~.
\end{equation}
are the reflection and transmission amplitudes for the ground state of
the atom. Note that all the probability amplitudes for the excited or
ground state of the atom depends on the coherent addition of amplitudes
of the barrier and well. We have recently shown that transmission of an
ultra-cold two-level atom in the excited state through two successive  
cavities depends strongly on this {\bf coherent} addition of amplitudes. 
This leads to the splitting of the transmission resonances of the
single cavity \cite{gsa}. 

\section{Phase time for ultra-cold atoms passing through a high quality
cavity - analog of sub - superluminal propagation}
In the previous section, we have seen that dynamics of an ultra-cold
atom passing through the cavity is reduced to the problem of reflection
and transmission of an atom incident on the cavity induced potentials. 
In this section, we study in detail the transmission of the atom in 
the initial excited state through the cavity initially in vacuum 
state. The transmitted part of the atom - field system is then 
given by using Eq. $(\ref{main2})$ 
\begin{equation}
|\Psi(z,t)\rangle = \int_{-\infty}^{\infty} dk~A(k)
~e^{-i \left(\hbar k^2/2 m \right) t}~T_{e,0}(k)~e^{i k z}~\theta(z-L)
~|e,0\rangle ~.
\label{main}
\end{equation}
The transmission amplitude $T_{e,0} \equiv |T_{e,0}| e^{i \phi(k)} $ given  
by Eq. $(\ref{excite})$ depends on the vacuum coupling energy $\hbar g$.
We consider a Guassian wave packet $A(k) = \exp\left({-(k - \bar{k})}^2 /
\sigma^2 \right)$ of width $\sigma$ and mean momentum $\bar{k}$ for the 
incident atom. With this substitution for $A(k)$, the wave function  
including the normalization factor, is given by 
\begin{equation}
|\Psi(z,t)\rangle = \frac{1}{{(2 \pi)}^{3/4}}~\sqrt{\frac{2}
{\sigma}} \int_{-\infty}^{\infty}~dk~\exp\left({-(k - \bar{k})}^2 /
\sigma^2 \right)~e^{-i \left(\hbar k^2/2 m \right) t}~|T_{e,0}| 
~e^{i \phi(k)}~e^{i k z}~|e,0\rangle~,~~~~~~z \ge L~.
\label{result}
\end{equation} 
For small width $\sigma$ the integrand in Eq. $(\ref{result})$ has
non vanishing value only in a small range of wave numbers $k$ 
centered about the mean $\bar{k}$. Then, the envelope of the 
transmitted wave packet ${|\langle e,0|\Psi(z,t)\rangle|}^2$ 
will be maximum when the total phase $\theta(k)$ of the integrand
exhibits extremum at the wave number $k = \bar{k}$. Since we have
assumed that the center of incident wave packet enters the cavity at 
time $t = 0$, this stationary phase condition gives the time at which
the wave packet at the exit of the cavity $z = L$, is peaked as 
follows :  
\begin{equation} 
\left. \frac{\partial \theta(k)}{\partial k}\right|_{k = \bar{k}}  
= \left. \frac{\partial}{\partial k} \left[k L + \phi(k) -
\left(\hbar k^2/2 m \right) t \right] \right|_{k = \bar{k}} = 0~,
\end{equation}
which yields the phase tunneling time $t_{ph}$  
\begin{equation}
t_{ph} = {\left[\frac{m}{\hbar k} \left(\frac{\partial \phi}
{\partial k} + L \right) \right]}_{k = \bar{k}}~. 
\label{phases}
\end{equation}
The integral in Eq. $(\ref{result})$ can be evaluated approximately
by making Taylor expansion of the phase of transmission amplitude
about the mean wave number $k = \bar{k}$. Keeping terms upto secend
order in the expansion and assuming $\sigma << \bar{k}$ to approximate
$|T_{e,0}(k)| \approx |T_{e,0}(\bar{k})|$, the transmitted wave
function is given at $z = L$ by 
\begin{eqnarray}
\left. |\Psi(z,t)\rangle~\right|_{z = L} &\approx& \frac{1}{{(2 \pi)}^{3/4}}~
\sqrt{\frac{2}{\sigma}}~\exp\left(i(\bar{k}L + \phi(\bar{k}) -
\bar{E} t/\hbar)\right)~|T_{e,0}(\bar{k})| \nonumber \\
&~~&~~~~~~~~\times \sqrt{\frac{2 \pi}{\left(\frac{2}{\sigma^2} + i \alpha 
\right)}} \exp\left(\frac{- \bar{E} (t - t_{ph})^{2}}{m \left(\frac{2}
{\sigma^2} + i \alpha  \right)}\right)~|e,0\rangle~,
\end{eqnarray}
where $\bar{E} = \hbar^2 \bar{k}^2 /2 m$ is the average energy of the 
incident atom and the parameter $\alpha = \frac{\hbar t}{m} - \left.
\frac{\partial^2\phi} {\partial k^2} \right|_{k = \bar{k}}$ accounts   
for the spreading of the wave packet as it propagates. The maximum 
amplitude of the transmitted wave packet occurs at time $t = t_{ph}$
given by the stationary phase assumption. It is very important to 
note that the phase time has no significance when either the Taylor 
expansion of the phase does not converge or additional terms
more than the second order term are important in the expansion.
In this general case, the transmitted wave packet will be deformed
from the Guassian shape and the concept of following the peak of the wave packet 
is meaningless.
When there is no cavity $|T_{e,0}(k)| = 1$,~$\phi(k) = 0$, then the
phase time in Eq. $(\ref{phases})$ becomes $t_{ph} = m L/\hbar \bar{k}
\equiv t_{cl}$ which is the classical time needed for the center of a free 
atomic wave packet to traverse a distance of length $L$. The phase
tunneling time which a particle takes to traverse a {\bf potential barrier},
has been studied extensively by Hartmann \cite{hartmann}. The tunneling time for
a barrier is less than the time a free particle takes to traverse the
same distance in free space. Here, we report such a superclassical
traversal of the ultra-cold atom through the vacuum induced potentials.
Note that the temperature of the atom will be in the range $10^{-7}$
to $10^{-8}$ K if the coupling constant $g$ is in the range of
100-10 kHz and if the mean momentum $\bar{k}/k_o = 0.1$.  
It should be borne in mind that both barrier and well contribute
to the traversal time of ultra-cold atoms. 
Using Eq. $(\ref{phases})$, we plot in Fig. 3 the phase time 
as a function of the mean wave number $\bar{k}$ for the length of the
cavity $k_o L = 10 \pi$. The important result here is that the phase time exhibits 
the resonant behaviour of transmission probability and that the phase time is
less than the classical time $t_{cl}$. In a different context
viz. in the tunneling time of electrons passing through a finite
superlattice, a similar resonant behaviour is found \cite{mexican}. Another  
remarkable behaviour of phase time is that it can even be {\bf negative}. Negative 
phase time implies that the peak of the transmitted wave packet emerges out 
even before the peak of the incident wave packet enters the interaction
region. This can be understood from the interference between the 
incident wave and the wave that is reflected at the end of the cavity.
From Eq. $(\ref{phases})$, we see that when the derivative of the 
phase of transmission amplitude is negative and its absolute value 
is greater than the length $L$ of the cavity, the phase time becomes 
negative. Put in another way, when the phase function $\phi(k) +
k L $ has negative slope, the phase time takes negative values. 
In Fig. 4, we show the phase time for the parameter
$k_o L = \pi/2$. It is seen from the graph that for ultra-cold  
atoms $(\bar{k}/k_o << 1)$ the phase time is negative. For fast atoms      
$(\bar{k}/k_o >> 1)$, the phase time approaches the classical 
time as the transmission probability becomes closer to unity.
The phase time being negative is very similar to the concept of negative
group velocity in the case of electro-magnetic pulse propagation.
Here, the variation of the refractive index of the medium with 
respect to the frequency has steep negative slope leading to
superluminal propagation \cite{wang}. To understand the negative
phase time, we have also plotted in Fig. 5 the phase function 
$\phi(k) + k L$. The graph shows the expected negative slope for
ultra-cold atoms. 
 
\section{time dependence of the wave packet for ultra-cold atoms}        
To study the behaviour of actual envelope of the wave function, we 
evaluate numerically the integral Eq. $(\ref{result})$ which
describes the propagation of a Guassian wave packet of an excited
atom through the vacuum induced potentials. Garrett and McCumber 
\cite{garrett} carried out a similar numerical integration for the 
electric field amplitude of a Guassian light pulse passing through
an anamolous dispersive medium. In Fig. 6(a), we show the numerical results 
for the normalized probability density ${|\langle e,0|\Psi(z,t)\rangle |}^2/\sigma$ 
at the exit of the cavity $z = L$ as function of the time for the  
parameters $k_o L = \pi/2$, $\sigma/k_o = 0.01$, $\bar{k}/k_o = 0.1$.  
The peak of the transmitted wave packet occurs at the time  
$t/t_{cl} \approx -0.98$ which matches with the phase time in the Fig. 4
for the parameter $\bar{k}/k_o = 0.1$. Thus the wave packet
appears to travel backwards in time in the sense of tracing the
locus of maximum amplitude. The peak of the transmitted wave 
packet is formed even before the peak of the incident wave 
packet enters the cavity. For comparison, we have also plotted 
the envelope of the wave packet which travels through the
same distance of length $L$ in free space. The peak of the
free wave packet occurs at the expected classical time.
From the graph, we see that for ultra-cold atoms $(\bar{k}/k_o << 1)$ 
the propagation of the atom through the cavity is faster
than through the free space. In Fig. 6(b), we plot the envelope 
of the wave function for the parameters $k_o L = \pi/2$,
$\sigma/k_o = 0.01$, $\bar{k}/k_o = 10$. For the case of
fast atoms $(\bar{k}/k_o >> 1)$ the transmitted wave packet 
has maximum amplitude at the classical time $(t/t_{cl} 
\approx 1)$ as expected from Fig. 4. Thus, the peak of 
the transmitted wave packet occurs at the instant given 
by the expression for phase time Eq. $(\ref{phases})$ even if
that instant is earlier than the instant at which incident
wave packet enters the cavity. While this is generally true
for a narrow momentum distribution characterized by $\sigma <<
\bar{k}$ of the incident atom, strong deformation of the incident
wave packet sometimes makes the phase time meaningless.  

\section{splitting of the wave packet}
We have so far considered only the propagation of the atomic wave
packet in the initial excited state. But in a high quality
the atom-field interaction leads to photon emission by the
excited atom. We can also study the behaviour of the transmitted
wave packet ${|\langle g,1|\Psi(z,t)\rangle|}^2$ for the ground
state of the atom using Eq. $(\ref{main2})$. For the parameters
of Fig. 6(a), the phase time for the ground state $t_{ph}/t_{cl} 
\approx 0.45$ is positive but still a superclassical time. 
Numerical integration also gives the same time delay for the
transmitted wave packet. In Fig. 7, we show the behaviour of
the phase time for the wave packet corresponding to transmitted
atom in the ground state. This behaviour is to be compared with 
that of the phase time for the transmission in the excited state 
(Fig. 4). The two phase times differ considerably for cold atoms.
Generally, the difference in phase times for the ground and excited
states of the atom results in {\bf splitting} of the incident wave 
packet into two in the total transmission. But for the parameters of 
Fig. 6(a), the total transmission is dominated by the contribution
from the ground state and hence the splitting is not seen. 
In Fig. 8, we plot the transmitted wave packet for the ground  
and excited states together for comparison.
The graph shows the time delay between the atoms exiting the cavity 
in excited and ground states. 

The {\bf splitting} of the incident 
wave packet can also occur for a different reason as shown in the
Fig. 9 for the parameters $k_o L = 10 \pi$, $\sigma/k_o = 0.5$,
$\bar{k}/k_o = 10$. It is seen that the probability density
is zero at the classical time. This can be understood from the
Rabi oscillations between the internal states of the fast atoms. 
For fast atoms $(k/k_o >> 1)$, the transmission amplitude can be
approximated as $T_{e,o}(k) \approx \exp(-i k L) (\exp(i k_0^{+} L)
+ \exp(i k_0^{-} L))/2 $ where $k_0^{\pm}$ are given by 
Eq. $(\ref{wave})$. The transmission probability ${|T_{e,0}|}^2$
exhibits oscillatory behaviour as a function of momentum $k$ of
the incident atom. Moreover, at the mean wave number corresponding to 
the classical time, the transmission amplitude $T_{e,0}(\bar{k})
\approx \cos(g t_{cl}) = 0$. Thus the {\bf correlation with the 
internal dynamics} (Rabi oscillations) of the atom leads to
the splitting of the incident wave packet of the external motion.
Obviously since the wave packet is deformed for these parameters,
the phase time $(t_{ph}/t_{cl} \approx -0.62)$ loses its physical 
significance and does not represent the peak to peak traversal time.  
 
\section{summary}
We have considered the propagation of a Guassian wave packet of an 
excited two-level atom through a high quality cavity which is 
initially empty. The atom-field interaction is equivalent to the
reflection and transmission of the wave packet through the 
potentials created by the dressed states. This is perhaps one of 
the rare examples in physics where the tunneling time would depend
on the coherent addition of transmission amplitudes through a
barrier and a well. The phase tunneling time can exhibit both
super- and sub- classical traversal behaviour. For certain set
of parameter, the phase tunneling time for cold atoms can even
be negative. All this can be understood in terms of the 
dispersion characteristics of the phase of the transmission 
amplitude and is analogous to the dispersion of the refractive
index which leads to super and subluminal propagation. Numerical
integration is performed to calculate the time dependent wave
packet at the exit of the cavity. In most cases the peak of the
transmitted wave packet occurs at the instant given by the 
expression for the phase time even when it is negative. Thus the
phase time can be considered as the approximate time required
for the atom to traverse the cavity. Further, we demonstrate how
the correlation between internal and external dynamics of the 
atom leads to the splitting of the wave packet, a feature which
is unique to the traversal of atoms through a cavity. Finally,
we note that the vacuum field for the initial state of the 
cavity does not limit the study of tunneling time of the atom.
In a general Fock state, the potential energy of atom-field
interaction with the cavity induced potentials will be different
from that of vacuum field. Still, we can redefine the atom-field
coupling constant of the interaction to include this change and
the superclassical tunneling of ultra-cold atoms is a common
feature for a general Fock state of the cavity field.

One of us (R. A) thanks Dr. Kulkarni for discussions on numerical
integration techniques.

\newpage
\vspace*{1.0 cm}
\begin{figure}[h]
\hspace*{3.8 cm}
\epsfxsize 2.6in
\epsfysize 1.1in
\epsfbox{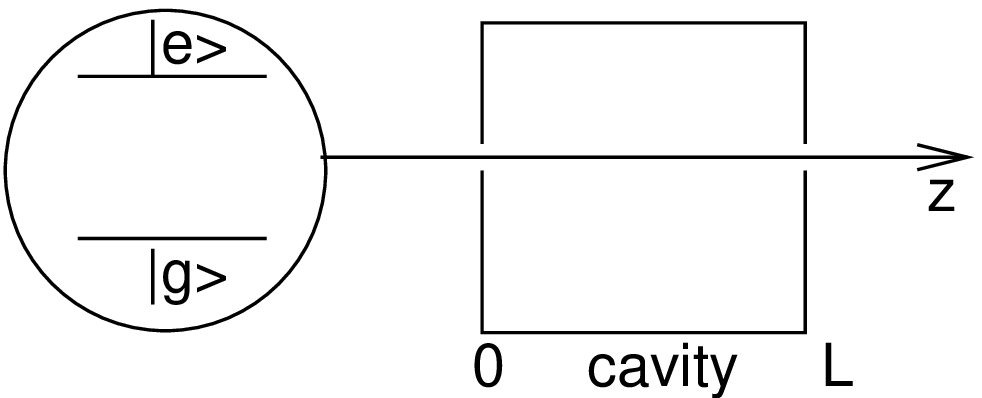}
\caption{The scheme of the high quality cavity with which the
ultra-cold atom interacts.}
\end{figure}

\vspace*{1.0 cm}
\begin{figure}[h]
\hspace*{3.0 cm}
\epsfxsize 3.2in
\epsfbox{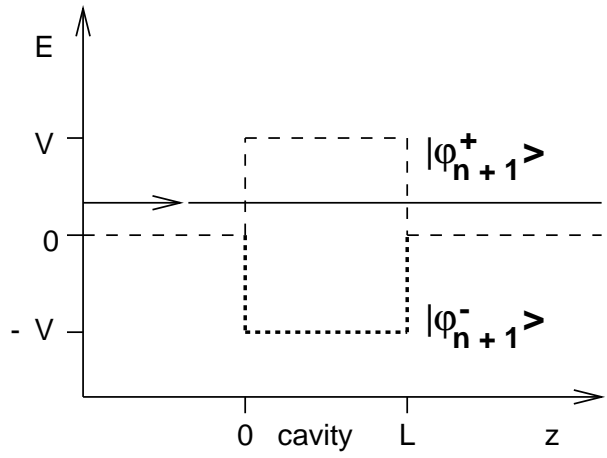}
\vspace*{0.4 cm}
\caption{Schematic representation of the energy E of the excited 
two-level atom incident upon a single mode cavity with n photons.
The interaction is equivalent to reflection and transmission of
the atom through a potential barrier (dashed) or potential well
(dotted) with a potential energy $V = \hbar g \sqrt{n + 1}$. The
atom can be reflected or transmitted in either of the states
$|e,n\rangle$ and $|g,n+1\rangle$.}
\end{figure}

\newpage
\begin{figure}[h]
\hspace*{1.5 cm}
\epsfxsize 4.5in
\epsfysize 3.0in
\epsfbox{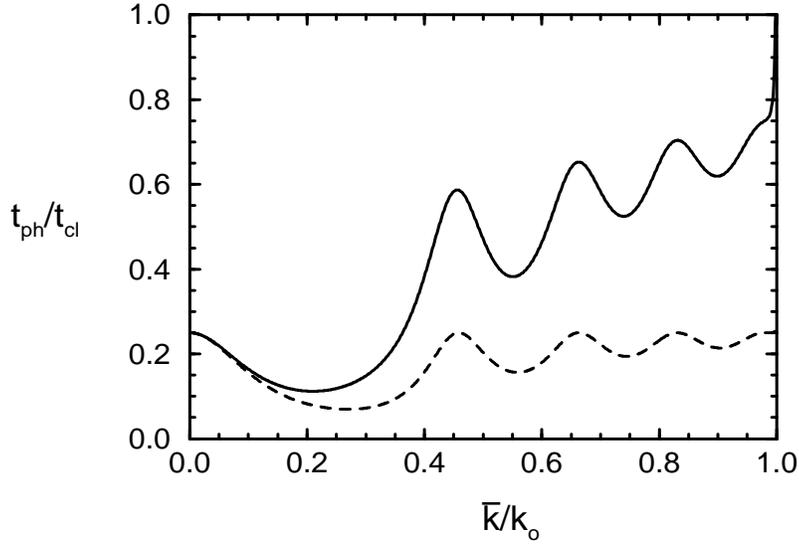}
\vspace*{0.4 cm}
\caption{The dependence of the dimensionless phase time (solid curve)
for transmission in the excited state on the mean wave number $\bar{k}/k_o$
of the incident atom for the parameter $k_o L = 10 \pi$. The phase time
follows the resonant behaviour of the transmission probability 
${|T_{e,0}|}^2$ (dashed curve).}
\end{figure}  

\begin{figure}[h]
\hspace*{1.5 cm}
\epsfxsize 4.5in
\epsfysize 3.0in
\epsfbox{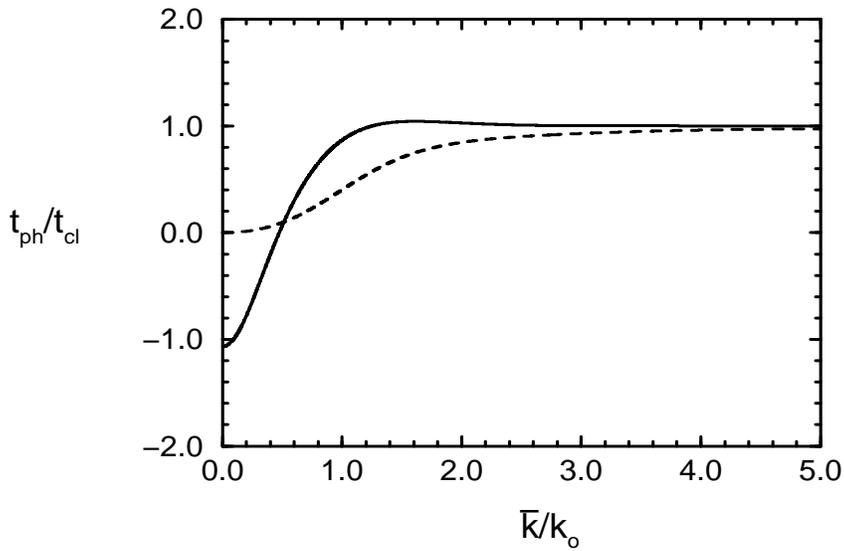}
\caption{The dimensionless phase time (solid curve) for transmission in the
excited state as a function of the mean wave number $\bar{k}/k_o$ of the
incident atom for the parameter $k_o L = \pi/2 $. The dashed curve 
represents the probability of transmission of the atom in the initial
excited state $({|T_{e,0}|}^2)$ through the cavity.}
\end{figure}

\newpage
\vspace*{-0.5 cm}
\begin{figure}[h]
\hspace*{1.5 cm}
\epsfxsize 4.5in
\epsfysize 3.0in
\epsfbox{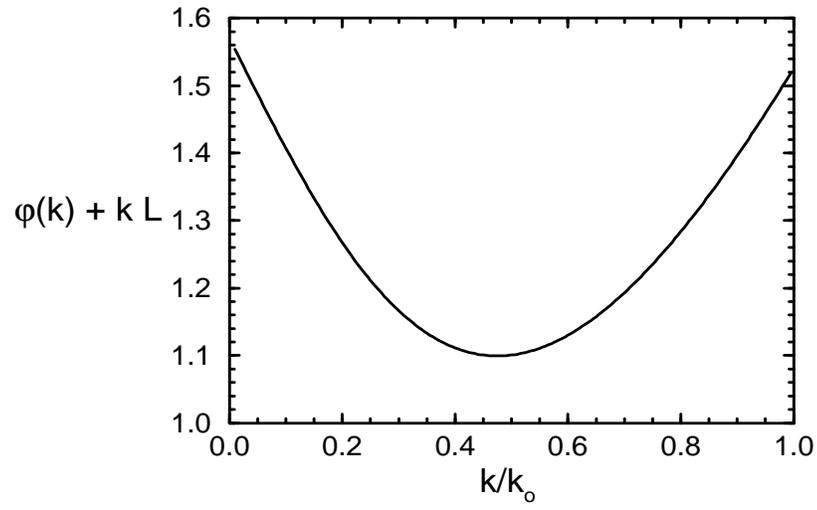}
\caption{The phase function $\phi + k L$ as a function of the wave
number $k/k_o$ of the excited atom for the parameter $k_o L = \pi/2 $.} 
\end{figure}

\newpage
\begin{figure}[h]
\hspace*{1.3 cm}
\epsfxsize=370pt
\epsfbox{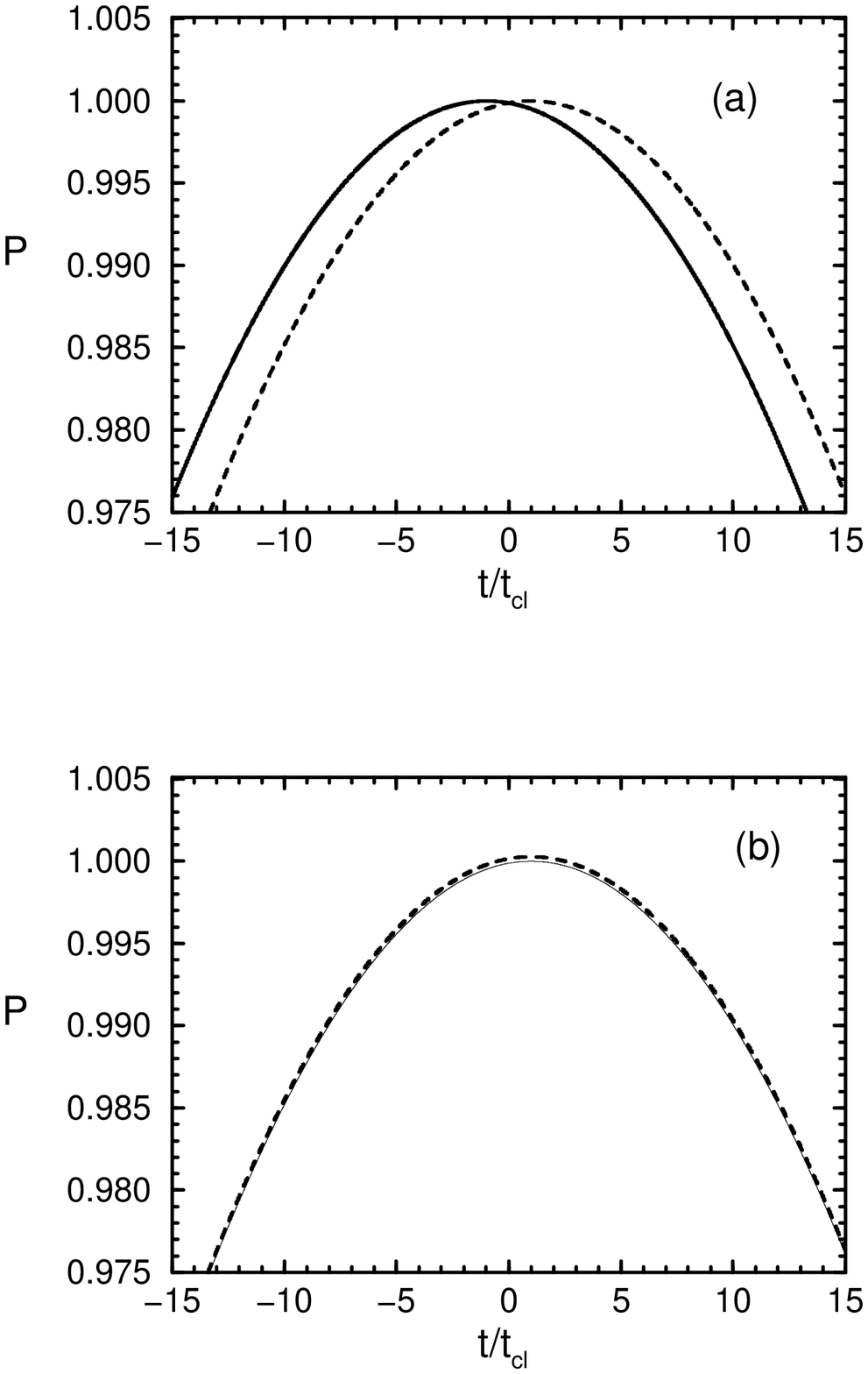}
\caption{The normalized probability density $P \equiv {|\langle e,0|\Psi(z,t)
\rangle|}^2/\sigma $ at $z = L$ as a function of dimensionless time $t/t_{cl}$.
The solid (dashed) curve represents $P$ after transmission through
the cavity (free space). The parameters used for the calculation
are $k_{o} L = \pi/2$, $\sigma/k_{o} = 0.01$ and (a) $\bar{k}/k_{o} =
0.1$, (b) $\bar{k}/k_{o} = 10$. Both the solid and dashed curves are
normalized to unity.}
\end{figure} 

\newpage
\begin{figure}[h]
\hspace*{1.5 cm}
\epsfxsize 4.5in
\epsfysize 3.0in
\epsfbox{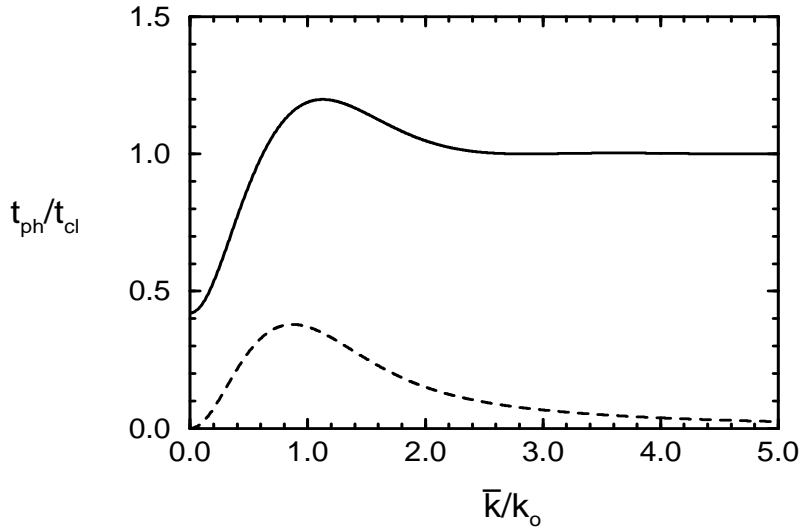}
\caption{The dimensionless phase time (solid curve) for transmission in the
ground state as a function of the mean wave number $\bar{k}/k_o$ of the 
incident atom for the parameter $k_o L = \pi/2 $. The dashed curve 
represents the probability of transmission of the atom in the ground state
$({|T_{g,1}|}^2)$ through the cavity.}
\end{figure}                            

\begin{figure}[h]
\hspace*{1.5 cm}
\epsfxsize 4.5in
\epsfysize 3.0in
\epsfbox{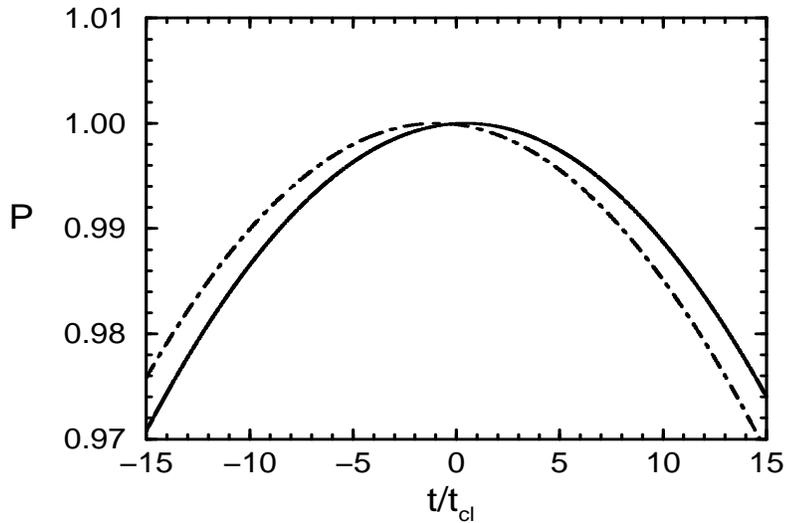}
\caption{The normalized probability density $P \equiv {|\langle g,1|\Psi(z,t)
\rangle|}^2/\sigma $ at $z = L$ as a function of dimensionless time $t/t_{cl}$.
The parameters of Fig. 6(a) are used for the calculation and dot-dashed 
curve corresponds to $P \equiv {|\langle e,0|\Psi(z,t)\rangle|}^2/\sigma$ 
at $z=L$. Both the solid and dot-dashed curves are normalized to unity.}
\end{figure}

\newpage
\vspace*{-0.5 cm}
\begin{figure}[h]
\hspace*{1.5 cm}
\epsfxsize 4.5in
\epsfysize 3.0in
\epsfbox{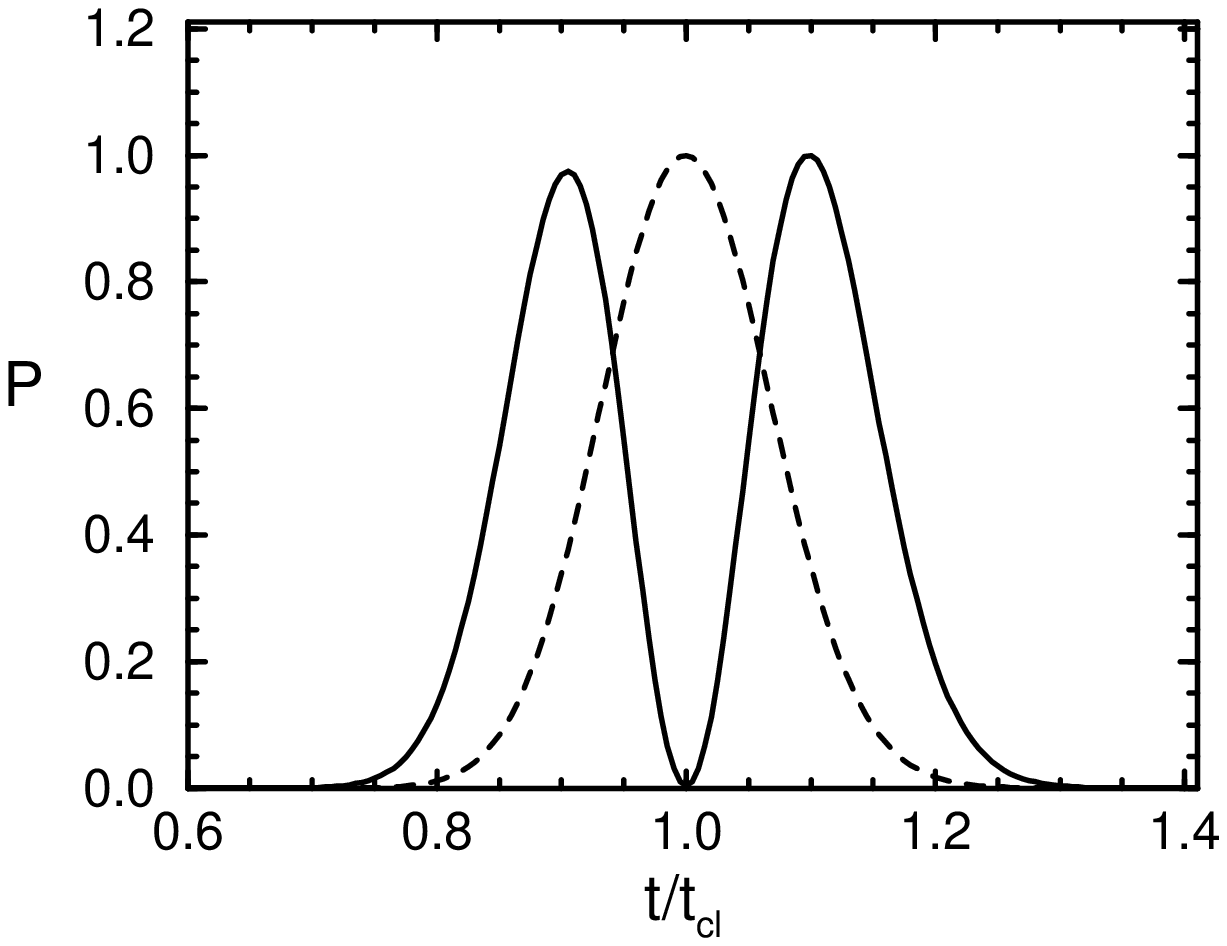}
\caption{The normalized probability density $P \equiv {|\langle e,0|\Psi(z,t)
\rangle|}^2/\sigma $ at $z = L$ as a function of dimensionless time $t/t_{cl}$.
The solid (dashed) curve represents $P$ after transmission through
the cavity (free space). The parameters used for the calculation are
$k_{o} L = 10 \pi$, $\sigma/k_{o} = 0.5$, and $\bar{k}/k_{o} = 10$.
Both the solid and dashed curves are normalized to unity.}
\end{figure}

\end{document}